\documentclass[superscriptaddress,twocolumn,showpacs,reprint,amsmath,amssymb]{revtex4}

\usepackage{graphicx}
\usepackage{epstopdf}
\usepackage{bm}
\usepackage{amsmath}

\newcommand{\ve}{\boldsymbol}
\newcommand{\tn}{\textnormal}
\newcommand{\re}{\textnormal{Re}\,}

\newcommand{\eq}{Eq.\,\eqref}

\begin{document}

\title{Ultrafast dynamics of fluctuations in \\ high-temperature superconductors far from equilibrium}

\author{L. Perfetti}
\affiliation{Laboratoire des Solides Irradi\'{e}s, Ecole polytechnique, 91128 Palaiseau cedex,
France}
\author{B. Sciolla}
\affiliation{Departement of Theoretical Physics, Ecole de Physique University of Geneva 24, Quai
Ernest Ansermet 1211 Gen\'eve, Switzerland}
\author{G. Biroli}
\affiliation{Institut de Physique Th\'eorique CEA, (CNRS URA 2306), 91191 Gif-sur-Yvette, France}
\author{C. J. van der Beek}
\affiliation{Laboratoire des Solides Irradi\'{e}s, Ecole polytechnique, 91128 Palaiseau cedex,
France}
\author{C. Piovera}
\affiliation{Laboratoire des Solides Irradi\'{e}s, Ecole polytechnique, 91128 Palaiseau cedex,
France}
\author{M. Wolf}
\affiliation{Fritz Haber Institute of the Max Planck Society, Faradayweg 4-6, 14195 Berlin,
Germany}
\author{T. Kampfrath}
\affiliation{Fritz Haber Institute of the Max Planck Society, Faradayweg 4-6, 14195 Berlin,
Germany}

\begin{abstract}

Despite extensive work on high-temperature superconductors, the critical behavior of an incipient
condensate has so far been studied exclusively under equilibrium conditions. Here, we excite
$\tn{Bi}_2\tn{Sr}_2\tn{CaCu}_2\tn{O}_{8+\delta}$ with a femtosecond laser pulse and monitor the
subsequent nonequilibrium dynamics of the mid-infrared conductivity. Our data allow us to
discriminate temperature regimes where superconductivity is either coherent, fluctuating or
vanishingly small. Above the transition temperature $T_\tn{c}$, we make the striking observation
that the relaxation to equilibrium exhibits power-law dynamics and scaling behavior, both for
optimally and underdoped superconductors. Our findings can in part be modeled using time-dependent
Ginzburg-Landau theory and provide strong indication of universality in systems far from
equilibrium.



\end{abstract}

\maketitle

Many physical properties of high-temperature superconductors cannot be understood in the framework
of a mean-field theory. For example, an incipient condensate with short coherence length fluctuates
in space and time, leading to precursor effects of the superconducting phase. As a consequence,
quantities such as conductivity~\cite{Alloul,Corson,Armitage,Bergeal}, heat
capacity~\cite{SpecificHeat}, diamagnetic susceptibility \cite{Ong1} and the Nernst signal
\cite{Ong2} may increase considerably when the temperature approaches the critical value
$T_\tn{c}$. For cuprates, most of these observations have successfully been described by the
Ginzburg-Landau (GL) model or quantum-field theories~\cite{Larkin,Tinkham,Cyrot,Gorkov}. Suitable
extensions of these concepts to systems in \emph{non}equilibrium states predict critical dynamics
and scaling laws~\cite{Polkovnikov}. Critical exponents independent of specific material properties
may help extend the idea of \lq\lq universality classes" \cite{Halperin} to systems far from
equilibrium. These phenomena should not only occur in condensed-matter systems but also in
high-energy physics and ultracold atomic gases~\cite{Polkovnikov,Lamporesi}.

An ideal protocol for the detection of critical dynamics requires that superconducting correlations
are monitored just above the transition temperature, where the system is expected to exhibit
universal behavior. To induce and probe the nonequilibrium regime, ultrafast pump-probe
spectroscopy is a very promising
approach~\cite{Demsar,Kabanov,Giannetti,Segre,Leitenstorfer,Averitt,Kaindl2,Perfetti,Cavalleri1,Madan,
Kaindl1,Kaindl3}. Unfortunately, in experiments using optical probes (photon energy of typically
1.5~eV), it is not straightforward to detect superconducting fluctuations above $T_\tn{c}$ because
of the presence of competing signals ascribed to the pseudogap~\cite{Demsar,Giannetti}. To
disentangle fluctuation- and pseudogap-related signal components, an optical three-pulse technique
was proposed recently~\cite{Madan}. Alternatively, electromagnetic pulses covering the
frequency range from about 1 to 3~THz have been shown to be highly sensitive probes of the
superconducting condensate~\cite{Averitt,Kaindl1,Kaindl2,Cavalleri1}. However, the duration of the
THz pulses ($\sim 0.5$ ps) is too long to resolve the ultrafast dynamics of the superconductor.

In this Letter, we monitor superconducting fluctuations by time-resolved detection of the
conductivity in the mid-infrared (MIR). Thin films of the superconductor
$\tn{Bi}_2\tn{Sr}_2\tn{CaCu}_2\tn{O}_{8+\delta}$ (BSCCO) are excited by optical laser pulses and
probed by electromagnetic pulses with center frequency of 20~THz and a duration of $\sim 100$~fs.
The corresponding photon energy of 80~meV is comparable to the single-particle gap of BSCCO,
thereby making MIR pulses a highly sensitive yet still ultrafast probe of superconducting
correlations~\cite{Leitenstorfer,Kaindl3}. By these means, we identify a critical scaling behavior
and a universal power law of the ultrafast nonequilibrium dynamics of the incipient condensate,
both for optimally and underdoped samples. Part of our data can be reproduced by time-dependent
Ginzburg-Landau theory, a coarse grained model justified near to the critical temperature. Our results suggest a promising route to study the highly unexplored field of critical behavior far from equilibrium.

\emph{Experimental details}.---Samples used are 150-nm thin films of optimally doped
($T_\tn{c}=91$~K) and underdoped ($T_\tn{c}=68$~K) BSCCO mounted on a diamond substrate that is
transparent to both pump and probe radiation. The MIR spectrometer is driven by laser pulses from a
Ti:sapphire laser oscillator (pulse duration of 10~fs, center wavelength of 780~nm, repetition rate
of 80~MHz). Part of the laser output is used to generate and detect MIR pulses~\cite{Leitenstorfer}
($\sim 100$~fs, spectrum from 10 to 30~THz), whereas the remainder is used to excite the sample
with an incident fluence of about $16~\mu\tn{J\,cm}^{-2}$. We simultaneously measure the MIR
electric fields $E_0(t,T)$ and $E(t,\tau)$ where $t$ denotes time with respect to the field maximum
at $t=0$ [see Fig. 1(a)], and $\tau$ is the delay between pump and probe pulse. While $E_0(t,T)$
quantifies the field transmitted through the sample in thermodynamic equilibrium at temperature
$T$, $E(t,\tau)$ is the field in case of the pumped sample.

\begin{figure} \begin{center}
\includegraphics[width=1\columnwidth]{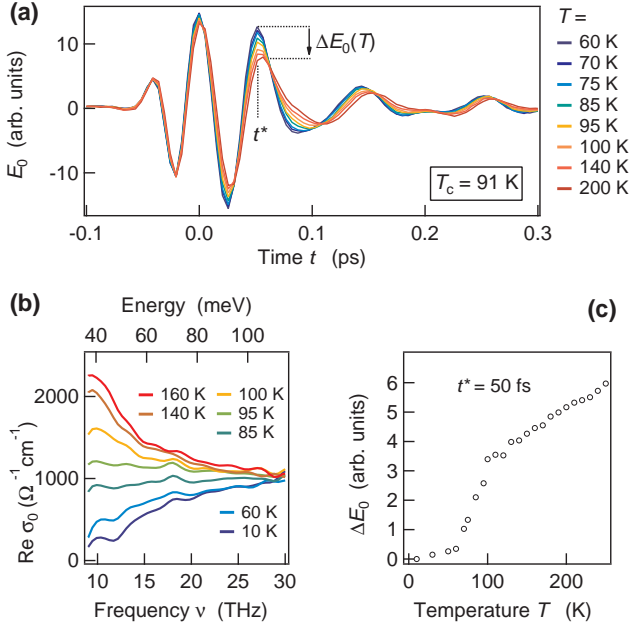}

\caption{Equilibrium data of optimally doped BSCCO ($T_\tn{c}=91$~K). (a)~Electric field $E_0(t,T)$
of the MIR waveform transmitted through the BSCCO film as a function of time $t$. Each sample
temperature $T$ is indicated by a different line color. (b)~Real part $\re\sigma_0(\nu)$ of the MIR
conductivity vs frequency $\nu$ and (c)~change $\Delta E_0(T)=E_0(t^*,6~\tn{K})-E_0(t^*,T)$ in the
transmitted MIR field at time $t^*=50$~fs as a function of $T$.} \label{Fig1}
\end{center}
\end{figure}

\emph{Equilibrium case}.---Figure 1(a) shows MIR field traces $E_0(t,T)$ after having traversed
optimally doped BSCCO films in equilibrium at various temperatures $T$. To extract the conductivity
$\sigma(\nu)$ as a function of MIR frequency $\nu$, we apply a Fourier transformation and standard
Fresnel formulas to the data of Fig. 1(a). The real part of the MIR conductivity is shown in Fig.
1(b), and we find that $\re\sigma(\nu)$ decreases with frequency for $T>T_\tn{c}=91$~K whereas it
follows the opposite trend in the superconducting phase with $T<T_\tn{c}$. These results are in
excellent agreement with previous reports based on Fourier transform spectroscopy~\cite{Quijada}
and ellipsometry~\cite{VDM,Basov}.

A detailed analysis of the probing process~\cite{SuppInf} shows that the variation $\Delta
E_0(T)=E_0(t^*,6\,\tn{K})-E_0(t^*,T)$ of the MIR electric field at the maximum position $t^*=50$~fs
is proportional to the real part of the difference $\sigma_0(\nu,6\,\tn{K})-\sigma_0(\nu,T)$
averaged over the frequency interval from 15 to 25~THz. Figure 1(c) displays the temperature
dependence of $\Delta E_0(T)$. When temperature is lowered from 300 to 100~K, we observe a linear
decrease, that turns into an abrupt drop at the transition to the superconducting phase ($T\sim
T_\tn{c}=91$~K). The physical interpretation of this sudden decrease is straightforward: the
emergence of the superconductivity gap ($T\sim T_\tn{c}$) reduces the density of electronic states
(DOS) in the vicinity of the Fermi energy, thereby reducing the number of scattering channels of
the electrons~\cite{Carbotte}. As a consequence, there is much less dissipation of the driving MIR
electric field below $T_\tn{c}$.

In the fluctuating regime above $T_\tn{c}$, the DOS becomes a function of local gap amplitude and
of the superconducting coherence length $\xi$~\cite{Carbotte}. The latter reflects the typical
distance over which the superconducting order parameter maintains its amplitude and
phase~\cite{Tinkham}, and it is expected to diverge exactly at the transition temperature.
Theoretical models of two-dimensional superconductors~\cite{Varlamov} predict that the MIR
conductivity scales with $\ln(\xi/\xi_0)$ where $\xi_0$ is the value of the coherence length at
zero temperature. Therefore, MIR pulses are an excellent probe of superconducting
fluctuations~\cite{Armitage}.

\begin{figure} \begin{center}
\includegraphics[width=1\columnwidth]{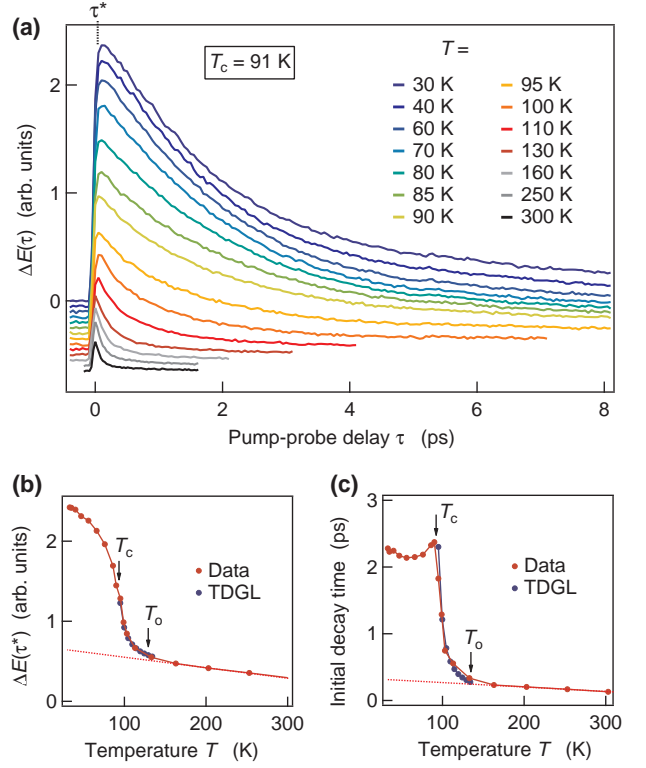}

\caption{Nonequilibrium data of optimally doped BSCCO ($T_\tn{c}=91$\,K). (a)~Pump-induced changes
$\Delta E(\tau)=E(t^*,-1\,\tn{ps})-E(t^*,\tau)$ in the transmitted MIR electric field as a function
of pump-probe delay $\tau$. $\Delta E(\tau)$ scales with the pump-induced change
$\re\Delta\sigma(\nu,T)$ in the instantaneous BSCCO conductivity averaged from 15 to 25~THz.
(b)~Signal magnitude $\Delta E$ and (c)~relaxation time $(\partial\ln\Delta E/\partial\tau)^{-1}$
directly after sample excitation ($\tau=\tau^*=50$~fs) as a function of temperature (red circles).
Blue squares result from a model based on the TDGL equation. The transition temperature
($T_\tn{c}=91$\,K) and the onset of superconducting fluctuations (at $T_\tn{o}=130$\,K) are
indicated by arrows.}\label{Fig.2}
\end{center}
\end{figure}

\emph{Nonequilibrium case}.---We now photoexcite the BSCCO sample with an optical pump pulse
(duration of 10\,fs, photon energy of 1.55\,eV) and monitor the resulting ultrafast dynamics.
According to previous work~\cite{Kaindl1,Lanzara}, our pump fluence of $16~\mu\tn{J\,cm}^{-2}$ is
large enough to reduce the order parameter by a fraction on the order of 80~\%. Figure 2(a) shows
the pump-induced change $\Delta E(\tau)=E(t^*,-1\,\tn{ps})-E(t^*,\tau)$ in the transmitted MIR
field for the optimally doped sample as a function of the pump-probe delay $\tau$ at various
ambient temperatures $T$. As above, we consider the MIR fields at time $t^*=50$~fs such that
$\Delta E(\tau)$ scales with the pump-induced change $\re\Delta\sigma(\nu,\tau)$ in the
instantaneous BSCCO conductivity averaged from 15 to 25~THz~\cite{SuppInf}. As shown by Fig. 2(a),
magnitude and recovery time of $\Delta E(\tau)$ strongly depend on sample temperature. Also, the
dynamics deviates substantially from an exponential decay and slows down with increasing pump-probe
delay.

To better illustrate the temperature dependence of the curves in Fig. 2(a), we extract two
characteristic parameters, namely the signal magnitude $\Delta E(\tau)$ and the instantaneous decay
time $(\partial\ln\Delta E/\partial\tau)^{-1}$ directly after sample excitation at
$\tau=\tau^*=50$~fs. The latter is obtained by fitting the dynamics with an exponential fit in a
small time interval just after photoexcitation. The two parameters are displayed as a function of
sample temperature $T$ in Figs. 2(b) and 2(c). As seen in Fig. 2(b), $\Delta E(\tau^*)$ is much
larger when $T$ is below the transition temperature. Above $T_\tn{c}$, we still observe an
appreciable $\Delta E(\tau^*)$ that above a temperature $T_\tn{o}$ turns into a relatively flat and
linearly decreasing curve (see arrows). Figure 2(c) shows that the initial decay time of the
pump-probe signal exhibits an analogous behavior: it is large below $T_\tn{c}$, then drops sharply
and turns into a flat curve above $T_\tn{o}$. We attribute these temperature intervals to the three
different regimes: a superconducting phase ($T<T_\tn{c}$), a fluctuating superconductor
($T_\tn{c}<T<T_\tn{o}$) and a metallic phase ($T>T_\tn{o}$). From Fig. 2, we obtain the rough
estimate $T_\tn{o}=130$\,K, which is in good agreement with values extracted from measurements of
transport~\cite{Alloul}, specific heat~\cite{SpecificHeat}, diamagnetic susceptibility~\cite{Ong1}
and the Nernst effect~\cite{Ong2}.


\emph{Ultrafast scaling behavior}.---In the temperature region from $T=T_\tn{o}$ down to the
critical temperature $T_\tn{c}$, the system is governed by an increasing coherence length
$\xi$~\cite{Tinkham}. The characteristic time $\tau_\tn{c}$ it takes a Cooper pair to cover the
coherence length should increase accordingly. The scaling hypothesis suggests that the relaxation
should follow a universal power law $[\tau/\tau_\tn{c}(T)]^{-\alpha}$ where $\alpha$ is a critical
exponent that only depends on the observable considered, the symmetry of the order parameter and
the dimension of the system~\cite{Bray}.

\begin{figure} \begin{center}
\includegraphics[width=1\columnwidth]{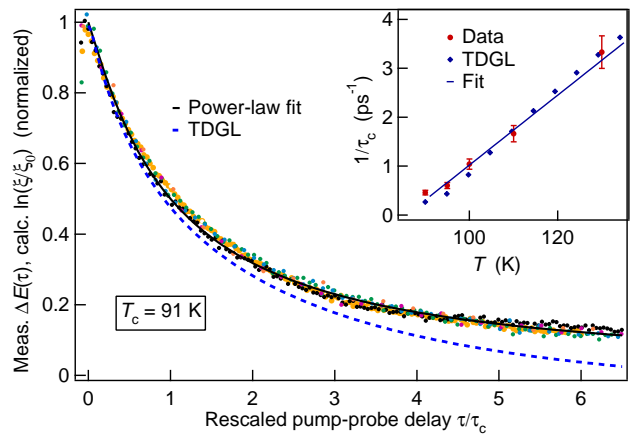}

\caption{Scaling behavior of ultrafast relaxation in the critical region between $T_\tn{c}$ and
$T_\tn{o}$ of optimally doped BSCCO ($T_\tn{c}=91$\,K). Measured pump-induced changes $\Delta
E(\tau)$ in the transmitted electric field as a function of rescaled pump-probe delay
$\tau/\tau_\tn{c}(T)$ at temperatures $T =90, 95, 100, 110,$ and 130\,K. Curves are also normalized
to the maximum signal value. The solid black line is a power-law fit [see \eq{eq:PL}] whereas the
dashed line displays the logarithm of the GL correlation length as calculated using the TDGL model.
The inset shows the scaling factor $\tau_\tn{c}^{-1}$ versus temperature as derived from the
experiment (red circles) and TDGL theory (blue marks) as well as a linear fit (solid line).}
\label{Fig3}
\end{center}
\end{figure}

Figure 3 shows the key result of this paper: the striking observation that the apparently diverse
dynamics of Fig. 2(a) indeed obey a scaling law. As seen in Fig. 3, all decay curves for
$T_\tn{c}<T<T_\tn{o}$ fall onto one when the time axis $\tau$ is normalized by a suitable time
$\tau_\tn{c}(T)$. More precisely, we find that the relation
\begin{equation}
\Delta E(\tau)=\frac{\Delta E(\tau^*)}{1+[\tau/\tau_\tn{c}(T)]^\alpha}\label{eq:PL}
\end{equation}
and $\alpha=1.1$ provide a very good description of our experimental data. The inverse scaling
factor $\tau_\tn{c}^{-1}$ is plotted as a function of temperature in the inset of Fig. 3(a). It
approximately follows a linear relation $0.45\,\tn{ps}^{-1} + (T/T_\tn{c}-1)6.5\,\tn{ps}^{-1}$. The
strong temperature dependence of $\tau_\tn{c}$ is consistent with a critical slowing down and
increasing coherence length of superconducting fluctuations when approaching the transition
temperature. We note that this is the classical regime of a second order phase transition and distinct from the quantum-critical fluctuations
that have been suggested to explain the \lq\lq strange metal\rq\rq~properties~\cite{VDM,Varma}. A critical scaling law of nonequilibrium fluctuations with $T\rightarrow T_\tn{c}$ is analyzed here for the first time. This finding goes substantially beyond equilibrium works~\cite{Alloul,Corson,Armitage,Bergeal,SpecificHeat,Ong1,Ong2,Larkin,Tinkham} and
reveals properties of the superconducting condensate far from equilibrium.

\emph{Theoretical model}.---We now simulate our experimental findings using the two-dimensional,
time-dependent Ginzburg-Landau (TDGL) model for the position~($\ve{x}$)- and
time~($\tau$)-dependent order parameter $\psi(\ve{x},\tau)$~\cite{Cyrot,SuppInf}. From a
microscopic perspective, $\psi$ is the local pair potential leading to the single-particle gap. The
TDGL model is justified for BSCCO due to the strong anisotropy of this layered compound. Temporal
variations of the vector potential are neglected due to the large ratio between London penetration
depth and coherence length~\cite{Du,Vidal}. In this coarse-grained description, the Ginzburg-Landau
(GL) time $\tau_\tn{GL}$ is the characteristic response time of the order parameter
$\psi(\ve{x},\tau)$. The remaining degrees of freedom are assumed to evolve much faster than $\psi$
and merely remain as noise $\eta(\ve{x},\tau)$ that affects the dynamics of $\psi$ as a source
term. We note that for our experimental conditions, the nonlinear term in the TDGL equation is very
small, and the fluctuations can be considered to a very good approximation as a free classic
field~\cite{SuppInf}.


We identify the photoexcited state $\psi(\ve{x},\tau^*)$ with a stationary solution of the TDGL
equation at the elevated temperature $T+\Delta T$. The value $\Delta T=17$~K is obtained by
comparing the transient $\Delta E(\tau^*)$ of Fig. 2(b) with the equilibrium $\Delta E_0(T)$ of
Fig. 1(c). In contrast, the fast degrees of freedom summarized in $\eta$ are assumed to return to
their equilibrium configuration characterized by temperature $T$ directly after excitation. This
sudden-quench hypothesis is motivated by the observation that uncondensed electrons in the normal
phase transfer the pump-deposited excess heat to the lattice within $\sim 100$~fs~\cite{Perfetti}.
We model $\eta$ by white noise with correlation function
$\langle\eta(\ve{x},\tau)\eta(\ve{x}',\tau')\rangle=2Sk_\tn{B}T\delta(\ve{x}-\ve{x}')\delta(\tau-\tau')$
where $S$ is the area of the sample film. Finally, for the GL time, we assume
$\tau_\tn{GL}=(T/T_\tn{c}-1)^{-1}0.25$\,ps, consistent with the estimate
$\tau_\tn{GL}=48\pi\sigma_\tn{DC}\lambda^2/c^2$ based on temperature-dependent measurements of
penetration depth $\lambda$ and normal-state DC conductivity
$\sigma_\tn{DC}$~\cite{Quijada,Vidal,Martin}. From our modeling, we calculate the superconducting
coherence length as a function of time $\tau$ since optical excitation~\cite{SuppInf}. To enable
comparison with the measured $\Delta E(\tau)$, we exploit the recent theoretical
prediction~\cite{Varlamov} that the MIR conductivity of two-dimensional superconductors scales with
$\ln(\xi/\xi_0)$ .

\begin{figure} \begin{center}
\includegraphics[width=\columnwidth]{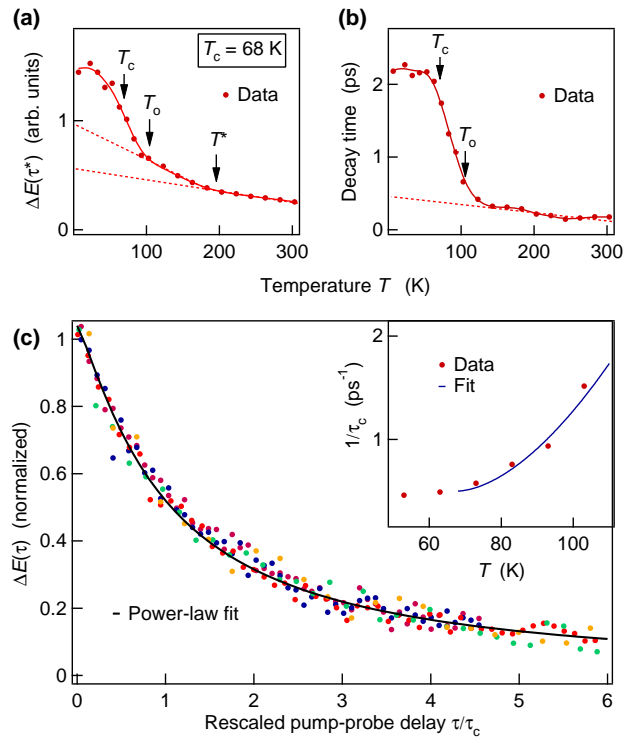}

\caption{Nonequilibrium data of underdoped BSCCO ($T_\tn{c}=68$~K) and their scaling behavior.
(a)~Magnitude of the pump-probe signal $\Delta E(\tau^*)$ and (b)~its relaxation time
$(\partial\ln\Delta E/\partial\tau)^{-1}$ right after sample excitation ($\tau^*=50$~fs) as a
function of temperature (red circles). Transition temperature $T_\tn{c}=68$~K, onset of
superconducting fluctuations $T_\tn{o}=100$~K and pseudogap temperature $T^*=200$\,K are indicated
by arrows. (c)~Measured pump-induced changes $\Delta E(\tau)$ of the transmitted electric field as
a function of rescaled pump-probe delay $\tau/\tau_\tn{c}(T)$ at temperatures $T =68, 78, 88, 98,
108$ and 130~K. Curves are also normalized to the maximum signal value. The solid black line is a
power-law fit [see \eq{eq:PL}]. The inset shows the scaling factor $\tau_\tn{c}^{-1}$ versus
temperature as derived from the experiment (red circles) and a fit (solid line, see text).}
\label{Fig4}
\end{center}
\end{figure}

\emph{Discussion}.---As shown by Figs. 2(b), 2(c) and 3, the calculated $\ln(\xi/\xi_0)$ matches
quite well the magnitude of the measured $\Delta E$ and the relaxation time for temperatures
between $T_\tn{c}+5~\tn{K}$ and $T_\tn{o}$. Below $T_\tn{c}+5~\tn{K}$, the TDGL overestimates
$\tau_\tn{c}$ by 70~\%. This discrepancy can be explained by an experimental crossover from the
sudden-quench regime to an adiabatic regime occurring below
$T_\tn{c}$~\cite{SuppInf,Tinkham,Polkovnikov}. In addition, the experimental curves of Fig. 3
follow a power law whereas the TDGL model predicts exponential relaxation. This discrepancy may
originate from a)~an incomplete thermalization of the fast degrees of freedom, thereby violating
the sudden-quench hypothesis, b)~the occurrence of coarsening phenomena related to extrinsic
disorder and c)~the signature of a conserved density coupled to the superconducting order
parameter. In case c), the TDGL (called model A in the Halperin classification scheme) would not be
adequate whereas an effective model belonging to a different Halperin class could give better
agreement with the experiment~\cite{Halperin}.

The limits of the TDGL description become more obvious in the critical dynamics of underdoped
BSCCO. Figure 4(a) shows the photoinduced change $\Delta E(\tau^*)$ in the transmitted probe field
measured for samples with $T_\tn{c}=68$~K. We identify the onset of superconducting fluctuations
with the upturn of the signal at $T_\tn{o}=100$~K. Note that $\Delta E(\tau^*)$ vs $T$ changes
slope at the temperature $T^*=200$~K. This kink is likely due to the progressive reduction of the
DOS known as \lq\lq pseudogap". The value of $T^*$ found here compares well to other experiments on
cuprates with similar doping level~\cite{Norman}. As shown by Fig. 4(b), the initial instantaneous
decay time $(\partial\ln\Delta E/\partial\tau)^{-1}$ displays a sudden upturn at the onset
temperature $T_\tn{o}=100$~K whereas it is insensitive to the crossover at $T^*$. In contrast to
the superconducting state, the pseudogap seems to lack any critical behavior. However, it is
possible that correlations responsible for the pseudogap formation are strongly coupled to Cooper
pairs and affect the dynamics of superconducting fluctuations near the critical point.

Figure 4(c) shows the relaxation dynamics of superconducting fluctuations in underdoped BSCCO.
Strikingly, as with optimally doped BSCCO, all rescaled curves in the critical region between
$T_\tn{c}$ and $T_\tn{o}$ are found to obey the power law of \eq{eq:PL}. The value $\alpha=1.2$
provides the best fit to the experimental data and is very similar to the $\alpha=1.1$ found for
optimally doped BSCCO (Fig. 3). The experimental evidence that the power law depends on the doping
level only weakly is another hint of universality under far-from-equilibrium conditions. In
contrast to optimally doped BSCCO (inset of Fig. 3), the temperature dependence of
$\tau_\tn{c}^{-1}$ is not linear, but approximately follows the function
$0.55\,\tn{ps}^{-1}+(T/T_\tn{c}-1)^\beta\,2.8\,\tn{ps}^{-1}$ with $\beta=1.7$ [see inset of Fig.
4(c)]. This finding is in contrast to the predictions of the TDGL and calls for extended
nonequilibrium models, perhaps including the coupling of the superconducting order parameter to
other conserved densities \cite{Halperin}.

In conclusion, we have observed universal behavior in the dynamics of superconducting fluctuations
in high-temperature superconductors far from equilibrium. The temporal relaxation of the MIR
conductivity follows the power law of \eq{eq:PL}, nearly independent of the doping level. We find
that the critical temporal scaling factor behaves as $(T-T_\tn{c})^\beta$ extending to temperatures
of up to $1.5T_\tn{c}$. The TDGL model reproduces some of the experimental results but suffers from
inherent limits. Our all-optical approach is applicable to micrometer-sized samples and opens new
perspectives in the field of critical systems far from equilibrium. Future work in this direction
will potentially identify the symmetry of a coarse-grained model describing high-temperature
superconductors~\cite{Halperin} and impose stringent constrains to microscopic theories.

We acknowledge Elio Tosatti for enlightening discussions on critical phenomena. This work is
supported by \lq\lq Investissements d'Avenir" LabEx PALM (ANR-10-LABX-0039-PALM) and the German
Research Foundation (Grant KA 3305/3-1).

\end{document}